\documentclass[twocolumn,showpacs,preprintnumbers,amsmath,amssymb]{revtex4}
\usepackage{graphicx}
\usepackage{dcolumn}
\usepackage{bm}

\begin{document}

\preprint{}

\title{Entanglement of the orbital angular momentum states of the photons
generated in a hot atomic ensemble}

\author{Qun-Feng Chen}
\author{Bao-Sen Shi}
\author{Yong-Sheng Zhang}
\author{Guang-Can Guo}
\affiliation{Key Laboratory of Quantum Information, University of Science and Technology of China, Hefei, 230026, People's Republic of China}

\date{\today}

\begin{abstract}
  Quantum protocols will be more efficient with high-dimensional entangled states.
  Photons carrying orbital angular momenta can be used to create a
  high-dimensional entangled state.
  In this paper we experimentally demonstrate the entanglement
  of the orbital
  angular momentum between the Stokes and anti-Stokes photons generated in a hot
  atomic ensemble using spontaneous four-wave-mixing. This experiment also
  suggests the existence of the entanglement concerned with spatial degrees
  of freedom between the hot atomic ensemble and the Stokes photon.
\end{abstract}


\maketitle

Entanglement is one of the most fantastic phenomenon of quantum mechanics,
and is used as a resource in quantum information field\cite{RevModPhys.74.347}.
High-dimensional two-particle entangled states can be used to realize
some
quantum information protocols more efficiently \cite{PhysRevA.64.012306, PhysRevLett.85.3313}.
Photons carrying orbital angular momenta (OAM) are used to create
high-dimensional entangled states, since OAM can be used to define an
infinite-dimensional Hilbert space \cite{calvo:013805}. The
first experiment of the entanglement of the OAM states
generated via spontaneous parametric down-conversion in a nonlinear
crystal was demonstrated in
2001\cite{Mair:N:2001:313}, since then
several protocols based on OAM states of photons have been
realized experimentally\cite{Vaziri:PRL:2002:240401, Vaziri:PRL:2003:227902,
Langford:PRL:2004:053601}. The transferring of OAM between classical light and cold
atoms\cite{PhysRevLett.83.4967, PhysRevLett.90.133001, Barreiro:OL:2004:1515}
and hot atoms\cite{Jiang:PRA:2006:043811} has also been reported in
the past years. 
Recently, the entanglement of OAM states of the photons generated in a
cold atomic system using
Duan-Lukin-Cirac-Zoller (DLCZ) scheme\cite{Duan:N:2001:413} has been
clarified by Inoue \emph{et al.}\cite{inoue:053809}. So far, there is no
experimental discussion about the entanglement of the OAM states of the
photons generated in a hot atomic system. In this paper we demonstrate
the entanglement of OAM states of the photons generated in a hot atomic ensemble
using the spontaneous four-wave-mixing (SFWM)\cite{balic:183601,Chen:unpublished}. Our
experiment is different from the experiment done by Inoue \emph{et al.}: In our experiment,
SFWM is used to generate a photon pair, in contrast with the experiment of
Ref.~\cite{inoue:053809}, in which the method based on DLCZ scheme is used. Furthermore,
our experiment is based on a hot atomic ensemble, which is more easy to be
realized compared with the scheme based on a cold atomci system. In our experiment, we
clearly demonstrate the entanglement of the OAM between the Stokes and
anti-Stokes photons generated via SFWM in a hot atomic ensemble, the
concurrence got in this experiment is about 0.81.
This experiment also suggests
the existence of the entanglement concerned with spatial degrees of freedom
between the hot atomic ensemble and the Stokes photon. 

\begin{figure}[tb]
  \begin{center}
	\includegraphics[width=8.3cm]{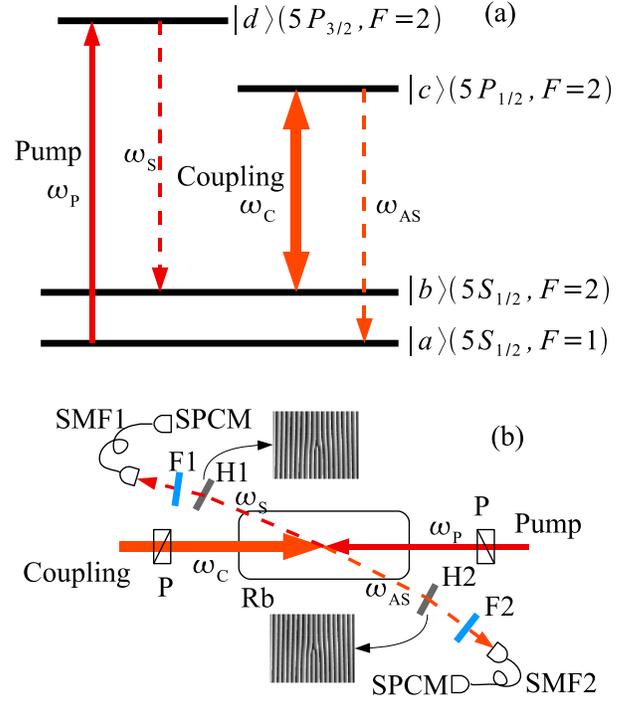}
  \end{center}
  \caption{(Color online) (a) Energy levels and frequencies of the
  lasers used in this experiment. (b) Schematic setup of our experiment. A strong coupling
  and a weak pump laser, which are resonant with $|5S_{1/2},F=2\rangle \to
|5P_{1/2},F=2\rangle$ and $|5S_{1/2},F=1\rangle \to
|5P_{3/2},F=2\rangle$ transitions of $^{87}$Rb respectively, are in
counter propagating. Pairs of correlated Stokes and anti-Stokes photons are generated
in phase-matched directions. H1 and H2 are computer-generated holograms; SMF1
and SMF2 are single-mode fibers, which are connected to single photon counting
modules(SPCM); F1 and F2 are filters.}
  \label{fig:1}
\end{figure}

The schematic setup used in this experiment is shown in Fig.~\ref{fig:1}. The energy
levels and the frequencies of the lasers used are shown in Fig.~\ref{fig:1}(a). A
strong coupling laser, which is resonant with the $|b\rangle\to|c\rangle$
transition, drives the populations of the atoms into level
$|a\rangle$. A weak pump laser, resonant with the
$|a\rangle\to|d\rangle$ transition, is applied to the system.  The
$|d\rangle\to|b\rangle$ transition will be induced by the pump laser
and the Stokes (S) photons will be generated. When a Stokes photon is
emitted, the atomic ensemble collapses into the state $\frac{1}{\sqrt{N}}
\sum_j|a_1,a_2,\ldots,b_j,\ldots,a_N\rangle$.  The strong coupling
laser repumps the atomic ensemble back to the state $|a_1,a_2,\ldots a_N\rangle$, and an
anti-Stokes (AS) photon is generated.  In this process, the energy, momentum and OAM of the
photons will be conserved\cite{Scully:PRL:2006:010501,Jiang:PRA:2006:043811},i.
e.,
\begin{eqnarray}
  \omega_{\rm S}+\omega_{\rm AS}&=&\omega_{\rm P}+\omega_{\rm
  C},\nonumber \\
  \vec k_{\rm S}+\vec k_{\rm AS}&=&\vec k_{\rm P}+\vec k_{\rm C},\nonumber\\
  L_{\rm S}+L_{\rm AS}&=&L_{\rm P}+L_{\rm C}\,,
  \label{cons}
\end{eqnarray}
where the $\omega_i$, $\vec k_i$ and $L_i$ represent the frequency, wave
vector and OAM of the corresponding photons respectively. According to Eq.~(\ref{cons}),
when the pump and coupling lasers carry zero OAM, the Stokes and
anti-Stokes photons will be in the entangled state of
\begin{equation}
  |\Psi\rangle= C\sum_{i=-\infty}^{+\infty}\alpha_i|i\rangle_{\rm
  S}|-i\rangle_{\rm AS}\,,
  \label{state}
\end{equation}
where $C$ is the normalization coefficient, $\alpha_i$ are the relative
amplitudes of the OAM states. In this work we only investigate the
entanglement concerned with $i=0$ and $1$, thus the experimental expected
entangled state can be written as:
\begin{equation}
  |\Psi\rangle = C(|0\rangle_{\rm S}|0\rangle_{\rm
  AS} + \alpha_1 |1\rangle_{\rm S}|-1\rangle_{\rm AS})\,.
  \label{}
\end{equation}
Although we only discuss the two dimensional case, it is natural to presume
that our discussion can be extended into high-dimensional cases over a wide
range of OAM\cite{inoue:053809}.

A Gaussian
mode beam carrying the well-defined OAM is in Laguerre-Gaussian (LG)
mode\cite{PhysRevA.45.8185}, it can be described by LG$_{pl}$ mode,
where
$p+1$ is the number of the radial nodes, and $l$ is the number of the
$2\pi$-phase variations along a closed path around the beam center. Here we only
consider the cases of $p=0$. The LG$_{0l}$ mode carries the corresponding OAM of
$l\hbar$ per photon and has a doughnut-shape intensity distribution:
\begin{equation}
  E_{0l}(r,\varphi) = E_{00}(r) \frac{1}{\sqrt{|l|!}}
  \left(\frac{r\sqrt{2}}{w}\right)^{|l|} e^{-il\varphi},
  \label{}
\end{equation}
where 
\begin{equation*}
E_{00}(r)= \sqrt{\frac{2}{\pi}}\frac{1}{w}\exp\left(-\frac{r^2}{w^2}\right)
\end{equation*}
is the intensity distribution of a Gaussian mode beam which carries zero OAM
(LG$_{00}$) and $w$ is the beam waist. In most cases, computer-generated
holograms
(CGH) are used to create the LG modes of various
orders\cite{Arlt:JOM:1998:1231}.
The superposition of the LG$_{00}$ mode and the LG$_{01}$ mode can be achieved by
shifting the dislocation of the hologram out of the beam center a
certain amount\cite{Mair:N:2001:313,Vaziri:JOO:2002:S47}.

In this paper, a CGH combined with a single-mode fiber are used for mode
discrimination. The $\pm 1$ order diffraction of the CGH increases the OAM of
the input beam by $\pm 1\hbar$ per photon when the dislocation of the hologram
is overlapped with the beam center. The first order diffraction of the CGH is
coupled into the single-mode fiber. The single-mode fiber collects only the
Gaussian mode beam, therefore the combination of the CGH and the single-mode
fiber can be used to select the LG$_{0\mp 1}$ or LG$_{00}$ mode or the
superposition of the them, according to which of the $\pm 1$ order
diffraction of the hologram is coupled and the displacement of the hologram. It should be
noted that there are also higher order LG modes in the first order diffraction,
but they are very small compared with the LG$_{0\pm1}$
mode\cite{Vaziri:JOO:2002:S47} and the influence of them is ignored in this paper.

The schematic experimental setup is shown in Fig.~\ref{fig:1}(b). A natural
rubidium cell with a length of 5 cm is used as the working medium. The
temperature of the cell is kept at about 50$^\circ$C, corresponding to an
atomic intensity of about $1\times10^{11}/{\rm cm}^3$.  The coupling laser,
which is vertically linear polarized, is resonant with the
$|5S_{1/2},F=2\rangle \to |5P_{1/2},F=2\rangle$ transition of $^{87}$Rb. The
intensity of the coupling laser is about 7 mW.  The pump laser, which is
counter-propagating with the coupling laser and horizontally polarized, is
resonant with the $|5S_{1/2},F=1\rangle \to |5P_{3/2},F=2\rangle$ transition of
${}^{87}$Rb.  The power of the pump is about $60\mu$W.  The $1/e^2$ diameters
of these two lasers are about 2 mm.  The vertically polarized Stokes photons
emitted at an angle of about 4$^\circ$ to the lasers are diffracted by a CGH
(H1), and the $-1$ order diffraction of the H1 is coupled into a single-mode
fiber (SMF1) after being filtered by the F1. The diffraction of the H1
decreases the OAM of the input photons by $1\hbar$ when the displacement of H1
is 0. The displacement of the CGH is defined as the distance between the
dislocation of the CGH and the beam center.  The horizontally polarized
anti-Stokes photons in the phase matched direction are diffracted by the other
CGH (H2). The $+1$ order diffraction is coupled into SMF2 after being filtered
by F2, which increases the OAM of the collected anti-Stokes photons by $1\hbar$
at 0 displacement. The diffraction efficiency of the CGHs used in this
experiment are about $40\%$.  Each of the filters F1 and F2 consists of an
optical pumped paraffin-coated $^{87}$Rb cell and a ruled diffraction grating.
The optical pumped rubidium cell is used to filter out the scattering of the
co-propagating laser, and the ruled diffraction grating is used to separate the
photons at the D1 and D2 transitions.  The collected photons are detected by
photon-counting modules (Perkin-Elmer SPCM-AQR-15).  The time resolved
coincident statistics of the Stokes and anti-Stokes photons are accumulated by
a time digitizer (FAST ComTec P7888-1E) with 2 n$s$ bin width and totally 160
bins. In this experiment the Stokes photons are used as the START of the
P7888-1E and the anti-Stokes photons after certain delay are used as the STOP
of the P7888-1E.

\begin{figure}[tb]
  \begin{center}
	\includegraphics[width=8.3cm]{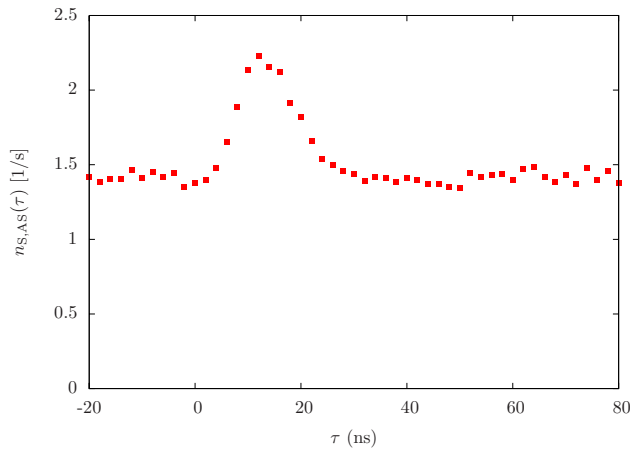}
  \end{center}
  \caption{(Color online) Time resolved coincidence counting between the Stokes and
  anti-Stokes photons. The data is accumulated about 1000 seconds and then
  normalized in time. $\tau$ is the relative delay between the Stokes and
  anti-Stokes photons.  The delay between the Stokes photons and anti-Stokes
  photons is cause by time used to generate anti-Stokes photons, which is
  mainly determined by the Rabi frequency of coupling
  field\cite{balic:183601}. 
  }
  \label{fig:2}
\end{figure}

The time resolved coincident counts of the Stokes and anti-Stokes photons when
the displacement of the both CGHs are far larger than the waists of the beam
are shown in Fig.~\ref{fig:2}. When the displacement of a CGH is far larger
than the waist of a beam, the CGH almost does not affect the mode of the
photons, therefore Fig.~\ref{fig:2} shows the coincidence between the Stokes
and anti-Stokes photons in LG$_{00}$ mode. The maximum coincident counts are
obtained at the relative delay of 12 ns between the Stokes and anti-Stokes
photons, which gives a correlation function of $g_{\rm S,AS}(12\textrm{
ns})=1.57\pm0.04$. The counting rates of the Stokes and anti-Stokes photons
are $1.4\times10^4/$s and $4.0\times10^4/$s respectively. The larger counting
rates of the anti-Stokes photons is caused by the atoms moving out and in the
coupling beam quickly, which makes a large effective decay rate between the
ground states. The atoms in the state $\left|b\right>$ moving into the
coupling laser contribute to uncorrelated anti-Stokes photons. Even when the
pump beam is absent the counting rate of the anti-Stokes is larger than
20000/s. These uncorrelated counts causes the large background in the
coincidence between the Stokes and anti-Stokes photons, as shown in
Fig.~\ref{fig:2}.  From Fig.~\ref{fig:2} we found that the correlated time
between the Stokes and anti-Stokes photons is less than 30 ns.

\begin{figure}[tb]
  \begin{center}
	\includegraphics[width=8.3cm]{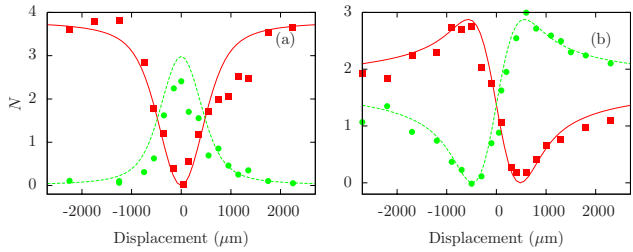}
  \end{center}
  \caption{(Color online) Coincident counts versus the displacement of H2 with
  different displacement of H1. (a) shows the results that the Stokes photons
  are in stationary states $|0\rangle$ (red squares) and
  $|1\rangle$ (green dots); (b) shows
  the results that the Stokes photons are in the superposition states
  $(|0\rangle\pm|1\rangle)/\sqrt{2}$. The data
  are fitted using the square of Eq.~(\ref{cocount}) with $w=0.8$ mm. }
  \label{fig:3}
\end{figure}

In order to evaluate the quantum correlation of the OAM states, we measure the
coincident counts with various displacements of the holograms.
Figure~\ref{fig:3} shows the results when the H1 is fixed at various
displacement while the displacement of H2 is swept. Every point is got by
$N=\sum_{\tau =2 \rm ns}^{32 \rm ns}(N(\tau)-bg)/bg$, where $N(\tau)$ is the
counting rate of each bin and $bg$ is background counting rate which is got by
averaging the coincidences between the Stokes and anti-Stokes photons when $\tau>50$
ns. This guarantees that most of the correlated anti-Stokes photons are taken
into account. Every point is accumulated over 500 seconds.  The data are fitted
with the square of the projection function\cite{Arlt:JOM:1998:1231}:
\begin{eqnarray}
  a(x0)&=&\int\!\!\!\int e^{-i
  \arg(r\cos\varphi-x0, r\sin\varphi)}\nonumber\\
  &&\times u_{AS}(r)u_{S}(r, \varphi)^*r\,\mathrm{d}
  r\,\mathrm{d}\varphi\,,
  \label{cocount}
\end{eqnarray}
where $\arg(x,y)$ is the argument of the complex number $x+i\,y$, $e^{-i
\arg(r\cos\varphi-x0, r\sin\varphi)}$ represents the transmitting function of
H2 with displacement of $x0$, $ u_{AS}(r)= E_{00}(r)$ is the field amplitude of
the anti-Stokes photons collected by the single-mode fiber after being
diffracted by the hologram,  and $u_{S}(r,\varphi) = \cos\theta
E_{00}(r)+\sin\theta E_{01}(r,\varphi)$ is the field amplitude of the Stokes photons collected by the
single-mode fiber. The superposition of the LG$_{00}$ and LG$_{01}$
modes can be controlled by the displacement of H1.
Equation~(\ref{cocount})
gives the projection between the different OAM modes. In this paper
the $u_{i}$s are the amplitudes of the Stokes and anti-Stokes
photons respectively. This equation is tenable only when the collapse of
the Stokes photons lead the anti-Stokes photons collapse into the
corresponding states. Therefore if
Eq.~(\ref{cocount}) always holds no matter the
Stokes photons are collapsed to stationary states or superposition states, the
Stokes photon and anti-Stokes photon should be in a quantum correlated state. In
Fig.~\ref{fig:3} (a), the red squares show the results of the
coincident counts versus the displacement of H2 when the displacement
of H1 is far larger than the waist of the Stokes photons, and the green
dots show the results when the displacement of H1 is 0.  The red line in
Fig.~\ref{fig:3} (a) is fitted with $\theta=0$ and the green dashed line
is fitted with $\theta=\pi/2$, which means the Stokes photons are in
LG$_{00}$ and LG$_{01}$ modes respectively. This figure demonstrates the
collapse of the Stokes photon state into the stationary states lead the anti-Stokes
photon state collapse into the corresponding stationary states. Therefore this figure indicates
clearly the correlation of OAM between the Stokes and anti-Stokes
photons. However, such a correlation can be obtained even in the mixture
$|0\rangle_{S}|0\rangle_{AS}$ and
$|1\rangle_{S}|1\rangle_{AS}$ states. To further demonstrate that the
Stokes and anti-Stokes photons are in a quantum correlated state, we displace the
H1 with a certain amount, which make the collected Stokes photons be
in the superposition states ${1}/{\sqrt{2}}(|0\rangle\pm |1\rangle)$,
and then sweep H2. The results are shown in Fig.~\ref{fig:3} (b). The data fit
well with the theoretical prediction, which demonstrates that the
anti-Stokes photon state collapses into the corresponding superposition
states when the Stokes photon
state
collapses into the superposition states.
Therefore the results shown in Fig.~\ref{fig:3} demonstrate that the Stokes
and anti-Stokes photons are in strongly quantum correlated OAM states.

\begin{figure}[tb]
  \begin{center}
	\includegraphics[width=6cm]{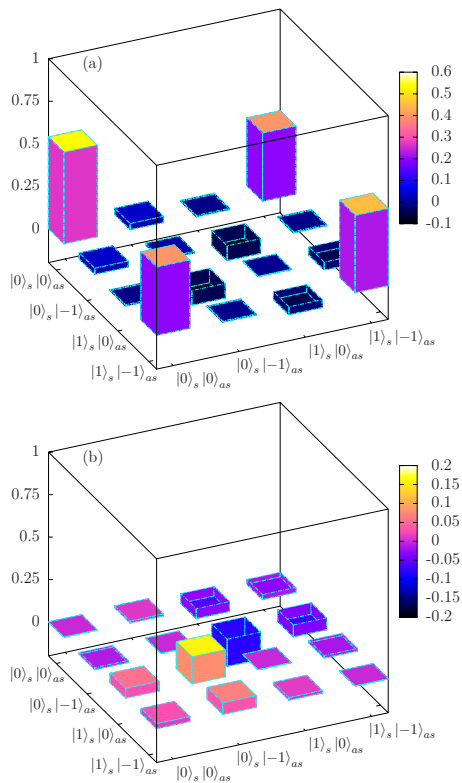}
  \end{center}
  \caption{(Color online) Graphical representation of the reconstructed density
  matrix. (a) is the real part and (b) is the imaginary part.}
  \label{fig:4}
\end{figure}

To further demonstrate the entanglement of the Stokes and anti-Stokes
photons, we perform a
two-qubit state tomography\cite{PhysRevA.64.052312}, and get the full
state of the Stokes and anti-Stokes photons. The density matrix is
reconstructed from the experimentally obtained coincidences with various
combinations of the measurement basis. A graphical representation of the
reconstructed density matrix is shown in Fig.~\ref{fig:4}. From the density
matrix, the fidelity\cite{Nielsen:2000} to the maximally entangled state $|\Psi\rangle =
(|0\rangle_{S}|0\rangle_{AS} +
|1\rangle_{S}|-1\rangle_{AS})/\sqrt{2}$ is estimated to about
$\langle \Psi|\rho|\Psi\rangle=0.89$. 
The concurrence\cite{PhysRevLett.80.2245} estimated from the density matrix is about $0.81>0$,
which demonstrated the Stokes and anti-Stokes photons are in an entangled
state clearly\cite{PhysRevLett.80.2245}.
The entanglement of formation\cite{Nielsen:2000} is also estimated to be 0.74.

The Stokes photons and the anti-Stokes photons are not generated
simultaneously in the SFWM. The atomic ensemble collapses into the state
$\frac{1}{\sqrt{N}} \sum_j|a_1,a_2,\ldots,b_j,\ldots,a_N\rangle$ after
emitting an Stokes photon, the information of the Stokes photons will be
stored in the atomic system firstly.  Lately the information of the atomic
ensemble is retrieved by the coupling laser, and an anti-Stokes photon is
generated\cite{Scully:PRL:2006:010501,inoue:053809}, the anti-Stokes photon
carries the information of the atomic ensemble. The speed of the anti-Stokes
photon generated is mainly determined by the Rabi frequency of the coupling
laser\cite{balic:183601}. Therefore the entanglement of OAM between Stokes
photons and the anti-Stokes photons might suggest the existence of the
entanglement of OAM between the Stokes photon and the atomic ensemble. Our
work is different from the work of V. Boyer \emph{el al.}\cite{boyer:143601}.
In their work they used a four-wave-mixing process\cite{boyer:143601:1} to
generate the spatially multimode quantum-correlated twin beams with finite OAM
in a hot atomic vapor. Their experiment is not a spontaneous process, and is
not in the photon level. They also have not demonstrated the entanglement
between the beams.

We estimate that the main sources of the errors in this experiment are from
follows: the decay rate of the atoms is very large, which causes the large
background counting; the instability of the frequency of the
lasers; there are also other LG modes in the diffraction except
for the LG$_{00}$, LG$_{01}$ modes and their
superposition\cite{Arlt:JOM:1998:1231,Vaziri:JOO:2002:S47}; the superposition
of the state LG$_{00}$ and LG$_{01}$ is got by shifting the hologram, which is
dependent on the beam waist, therefore the small fluctuation of the beam
position also causes the error. 

In summary, we have demonstrated the entanglement of OAM states between the
Stokes and anti-Stokes photons generated via SFWM in a hot rubidium cell. The
entanglement of the Stokes and anti-Stokes photons also suggests that the
Stokes photon might entangle with the hot atomic ensemble in spatial degrees
of freedom (OAM in this paper).

\begin{acknowledgments}
  We thank Pei Zhang for supplying computer-generated holograms and some
  useful discussion. We also thank Xi-Feng Ren for some useful discussion.
  This work is supported by National Fundamental Research
  Program(2006CB921907), National Natural Science Foundation of
  China(60621064, 10674126, 10674127), the Innovation funds from Chinese
  Academy of Sciences, and the Program for NCET.
\end{acknowledgments}


\end{document}